\newcommand{\p}	{{\partial}}

\documentclass{nature}

\usepackage{graphicx}
\makeatletter
\let\saved@includegraphics\includegraphics
\AtBeginDocument{\let\includegraphics\saved@includegraphics}
\renewenvironment*{figure}{\@float{figure}}{\end@float}
\makeatother

\usepackage{pdfpages}

\usepackage{tikz}

\title{Turbulent Action at a Distance due to Stellar Feedback in Magnetized Clouds }

\author{Stella S. R. Offner$^{1}$ \& Yue Liu$^2$}

\newcommand{\kms}	{{\rm km}\, {\rm s}^{-1}}
\def\rsun{\ifmmode {\rm R}_{\mathord\odot}\else $R_{\mathord\odot}$\fi}
\def\msun{\ifmmode {\rm M}_{\mathord\odot}\else $M_{\mathord\odot}$\fi}
\def\lsun{\ifmmode {\rm L}_{\mathord\odot}\else $L_{\mathord\odot}$\fi}

\begin{document}

\maketitle

\begin{affiliations}
 \item Department of Astronomy, The University of Texas at Austin, Austin, TX 78723
 \item Department of Astronomy, University of Massachusetts, Amherst, MA 01003 
\end{affiliations}

\begin{abstract}
A fundamental property of molecular clouds is that they are turbulent\cite{MandO07}, but how this turbulence is generated and maintained is unknown. One possibility is that stars forming within the cloud regenerate turbulence via their outflows, winds and radiation (``feedback")\cite{krumholz14ppvi}. Disentangling motions created by feedback from the initial cloud turbulence is challenging, however. Here we confront the relationship between stellar feedback and turbulence by identifying and separating the local and global impact of stellar winds. We analyze magnetohydrodyanamic simulations in which we track wind material as it interacts with the ambient cloud. By distinguishing between launched material, gas entrained by the wind and pristine gas we show energy is transferred away from the sources via magnetic waves excited by the expanding wind shells. This action at a distance enhances the fraction of stirring motion compared to compressing motion and produces a flatter velocity power spectrum. We conclude stellar feedback accounts for significant energy transfer within molecular clouds, an impact enhanced by magnetic waves, which have previously been neglected by observations. Altogether, stellar feedback can partially offset global turbulence dissipation. 
\end{abstract}

Supersonic turbulence shapes molecular cloud evolution, lifetimes, star formation efficiency, and even, the properties of the stars that form\cite{Offner14ppvi}. Molecular clouds likely inherit their initial turbulence from the interstellar medium (ISM), but these motions are expected to decay quickly\cite{stone98}. 
Several ideas have been proposed to explain observed cloud energetics. Cloud lifetimes may be short compared to the decay time\cite{elmegreen00}. 
Gravitational collapse may drive turbulence\cite{vazquez07,robertson12} or simply dominate the observed motions\cite{heitsch08}.  Alternatively, energy injected by forming stars may sustain turbulence\cite{krumholz14ppvi}. On parsec scales, both observational and numerical studies often find the energy associated with protostellar outflows, winds and radiation is comparable to the cloud energy\cite{krumholz14ppvi}. 
However, whether feedback from young sources contributes to the global cloud turbulent cascade or mainly affects nearby gas has not been demonstrated.  On galactic scales, winds from massive stars and supernovae are instrumental in regulating the ISM and total star formation rate\cite{hopkins14,agertz15,gatto17}. However, like the situation at smaller scales, a direct connection between global star formation rate and ISM turbulence is difficult to prove and could be explained by gravity\cite{caldu13,krumholz16}.  

Various numerical studies have concluded that stellar feedback can maintain turbulence on parsec scales when star formation is vigorous\cite{krumholz14ppvi,Offner15}.
However, none of these studies distinguished between the feedback itself, gas entrained by it and more removed, non-interacting gas. 
 If feedback energy is deposited locally, at least one dynamical time is required to distribute the energy throughout the cloud given feedback velocities\cite{matzner02} or it may not impact global energetics at all. A possible solution is provided by the magnetization of 
molecular clouds, where feedback energy could by carried away from sources by magnetosonic waves excited by outflow cavities or expanding shells
\cite{heitsch02,wang10,Offner17}. Numerical simulations have demonstrated that this mechanism can significantly enhance the energy of material in the vicinity of protostellar outflows on scales of $\sim 0.1$~pc\cite{Offner17}.  Directly observing the propagation of such waves is challenging.  In quiescent clouds, regularly spaced striations may be evidence of magnetosonic waves excited on super-cloud scales\cite{tritsis18}. There are hints that gas near protostellar outflows has larger linewidths and exhibits enhanced dissipation\cite{larson15}.  However, star-forming clouds are too kinematically complex to directly identify the propagation of coherent waves.

Several numerical studies have demonstrated feedback can leave an imprint in the gas density, velocity and intensity distributions, which is revealed quantitatively by turbulent statistics such as the power spectrum, bicoherence and spectral correlation function\cite{carroll09,hansen12,Offner15,boyden16}. 
The momentum power spectrum of NGC1333, a young region densely populated by outflows, exhibits a ``knee," which occurs at the characteristic scale of outflows and arguably attests to their impact\cite{swift08}. However, this has not been replicated elsewhere. Clear evidence of the impact of feedback on cloud turbulence may be absent because stellar feedback is not as important as simulations suggest or simply because signatures are difficult to cleanly identify given optical depth effects and variations in physical and chemical properties.  

 In this work we focus on stellar winds, because their impact is an order of magnitude larger than that of protostellar outflows, and their energy input is comparable to the cloud energy but is not sufficient to disperse the cloud\cite{arce11}.
We analyze a set of magnetohydrodynamic (MHD) simulations of  stellar sources embedded in turbulent clouds (see Methods). These stars launch winds according to a model for radiatively driven winds from main sequence stars\cite{Offner15}. 
The simulations are carried out with  {\sc ORION2}  and model a 5 pc piece of a turbulent molecular cloud with initial gas temperature 10~K, gas density $2\times 10^{-21}$ g cm$^{-3}$ and total mass 3,762 $\msun$. 
We randomly insert five stellar sources, which represent either young massive stars recently formed within the cloud or older stars that formed in a neighboring cloud and wandered from their birth site. The simulations follow two different stellar distributions (W1 and W2), turbulence patterns with four different magnetic field strengths (T0, T2, T3 and T4) and calculations with and without gravity (denoted with a G). 
Runs with no wind sources in which the initial turbulence naturally decays, runs with no initial turbulence (denoted with a B) and a run with $\mathbf{B}=0$ (T0) serve as baselines for the evolution in the absence of feedback, turbulence and magnetic fields, respectively. 
We exploit the multi-fluid functionality of {\sc ORION2} to track wind material\cite{Offner17}. This allows us to separate launched and entrained gas from pristine gas that has not directly interacted with the winds. We consider gas pristine if less than 1\% of the gas in the cell is wind material (i.e., $F_t < 0.01$, see Methods).   Full simulation details are given in the Methods section, and Supplementary Table 1 summarizes the calculation properties. 

\begin{figure}
\begin{center}
\includegraphics[scale=0.55]{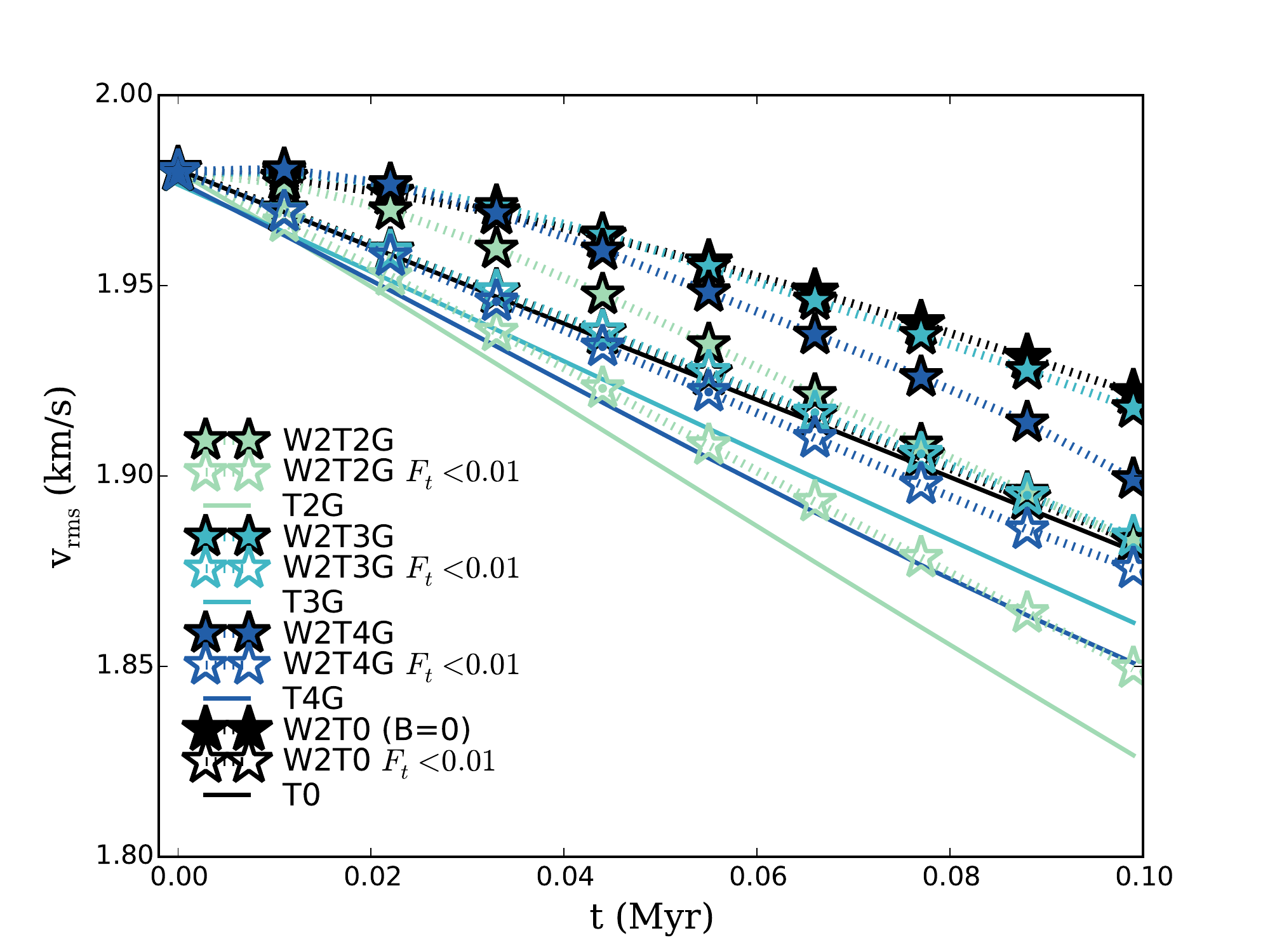}
\end{center}
\caption{Velocity dispersion, $v_{\rm rms}$, as a function of time, $t$. The filled stars indicate the total gas dispersion, including wind material. The empty stars indicate the velocity dispersion of pristine gas ($F_t < 0.01$). The solid lines show the velocity dispersion of gas in simulations without active winds. The magnitude of the difference between the open stars and the solid lines reveals the significance of the non-local driving. No offset appears in the non-magnetic run (black stars). 
\label{vdisp} }
\end{figure}

\begin{figure}
\begin{center}
\includegraphics[scale=0.6]{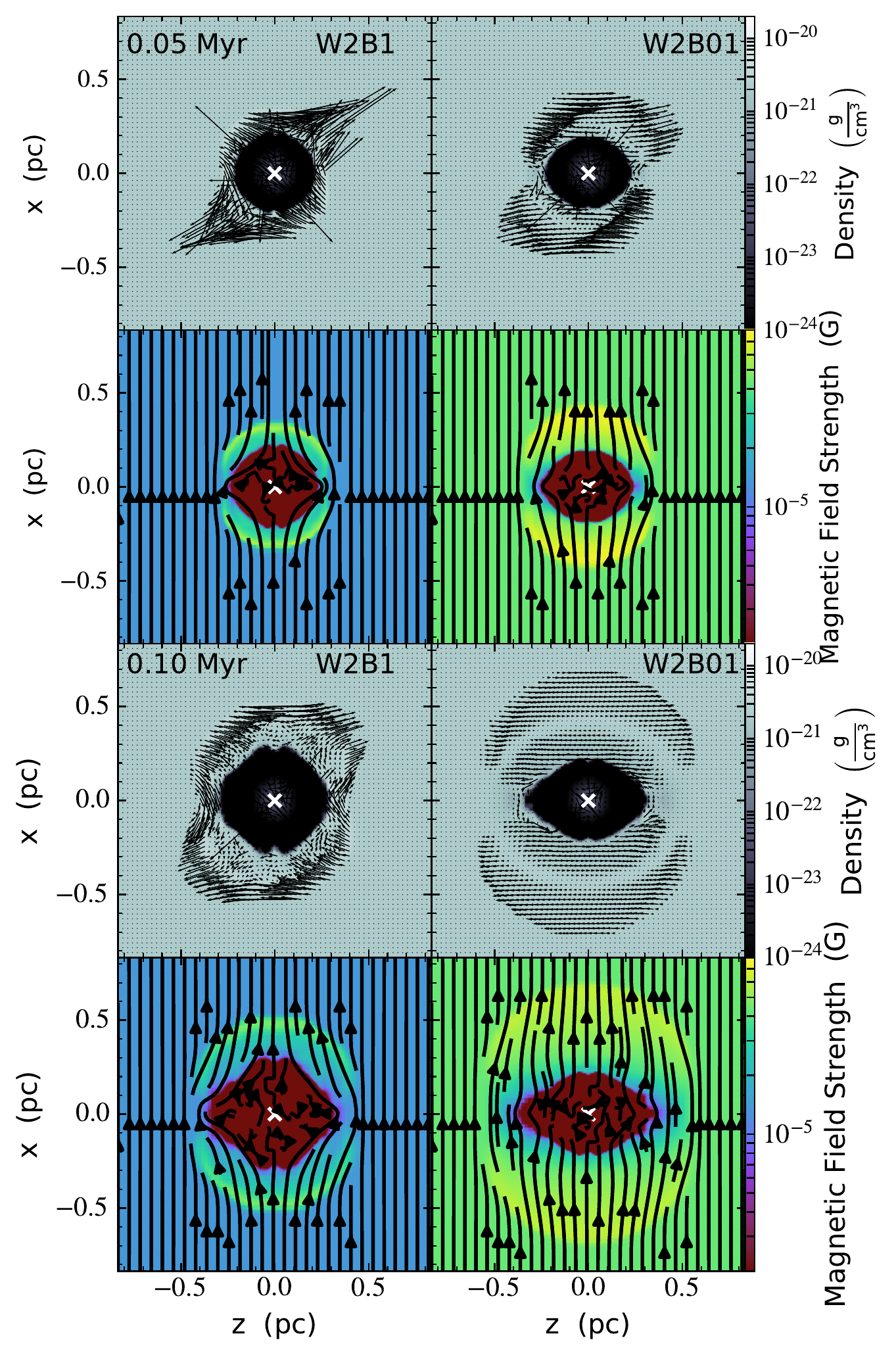}
\end{center}
\caption{Gas density and magnetic field strength ( $\mathbf{B}_{\rm rms}=\left(B_x^2+B_y^2+B_z^2 \right)^{1/2}$). Results for the non-turbulent calculations W2B1 (left panels) and W2B01 (right panels) at two different times. Arrows indicate the direction and relative magnitude of the gas momentum. Streamlines show the magnetic field lines. \label{bvst} }
\end{figure}

The gas velocity dispersion, as observationally measured using the line width, is the simplest metric of turbulence. Figure \ref{vdisp} shows the mass-weighted dispersion of the gas velocities as a function of time.   
Winds contribute significant energy, elevating the global velocity dispersion and reducing the impact of turbulent dissipation\cite{Offner15}.  The velocity dispersion for magnetized gas without wind material (open stars) is {\it higher} than the  velocity dispersion in runs without feedback. This offset does not occur for the run without magnetic fields (black), where the velocity dispersion of the pristine gas is the same as that in the non-feedback calculation.
The enhancement occurs because the expanding wind bubbles ``pluck" the field lines, exciting magnetosonic waves that propagate ahead of the shock front as illustrated in Figure \ref{bvst} and the supplementary movie.  
The non-uniform densities and magnetic field seeded by the turbulence produce conditions where the local Alfv\'en velocity $v_A =  B_{\rm rms}/\sqrt{4 \pi \rho} \sim 0.1-10^2~\kms$  (see supplementary material).  This indicates that while waves perturbed by bubble expansion may not reach all points on the domain within 0.1 Myr they do propagate significantly beyond the wind shock. 

To quantify the region of influence  we compare outputs from calculations with (W2T2) and without (T2) winds. After 0.1 Myr the volume filling fraction of the wind material is 1.6\% as defined by $F_t>0.01$. At the same time, 14\% of the W2T2 domain has a magnetic field strength that differs from that in the same location in T2 by more than 25\%, while 4\% of the volume differs by more than 50\%.  Alternatively, one can consider the significance of the effect via the fraction of the domain in which Alfv\'en waves can propagate faster than the initial turbulent velocity dispersion ($\sigma_{\rm 1D}=1.1~\kms$). In W2T2, 53\% of the volume has a local Alfv\'en velocity greater than $3~\kms$, while 9\% of the volume has an Alfv\'en velocity greater than 10~$\kms$ (see supplementary information). Thus,  Alfv\'en and fast magnetosonic waves can outpace typical turbulent motions in a large portion of the domain; the sphere of influence is significantly larger than the volume of the wind material. While waves excited by the expanding shells do not reach most of the domain during our calculation, the crossing time of waves traveling at 3 $\kms$ is comparable to the global gravitational freefall time, $\sim$1.4 Myr, where the turbulence crossing time $t\sim L/\sigma_{\rm 1D} \sim 4.5$ Myr given a cloud size of $L=5$ pc. This comparison underscores that waves excited by feedback traverse most of the cloud in a dynamical time. 

All the calculations, which have mean field strengths spanning $10 - 40 ~\mu$G show magnetic wave excitation, suggesting this mechanism may enhance turbulence even in weakly magnetized clouds.
We also expect turbulence enhancement in regions with higher mass ionizing sources, since magnetosonic waves outside expanding HII regions have previously been found in non-turbulent calculations\cite{gendelev12}.   
The simulations with and without self-gravity show nearly identical behavior.

Next, we consider the significance of the non-local energy deposition. The total energy injection rate, i.e., wind luminosity, is
$\dot E_{\rm W2} = \frac{1}{2} \dot M v_w^2 \simeq 14.6 L_\odot$,
and $\dot E_{\rm W1}  = 45.6 L_\odot$.   However, most of this energy is dissipated in the wind shock.  The global rate of energy loss due to shock dissipation in T2 and T2G is $\dot E = -1.5 L_\odot$ (see Supplementary Table 2). In comparison the energy loss for W2T2 and W2T2G excluding the wind material is $\dot E = -1.3 L_\odot$, while $\dot E = -1.1 L_\odot$ for W1T2.  This corresponds to a difference of $0.2-0.4 L_\odot$. While this is only 1\% of the wind input, it is sufficient to offset the global energy loss rate by $\sim 10-30$\%.  Including the wind contribution, the energy loss rate of W2T2 is $\dot E = -0.6 L_\odot$, 
so this means ``action at a distance" accounts for 
$\sim 40$\% of the total energy deposited by winds. 

\begin{figure}
\begin{center}
\includegraphics[scale=0.55]{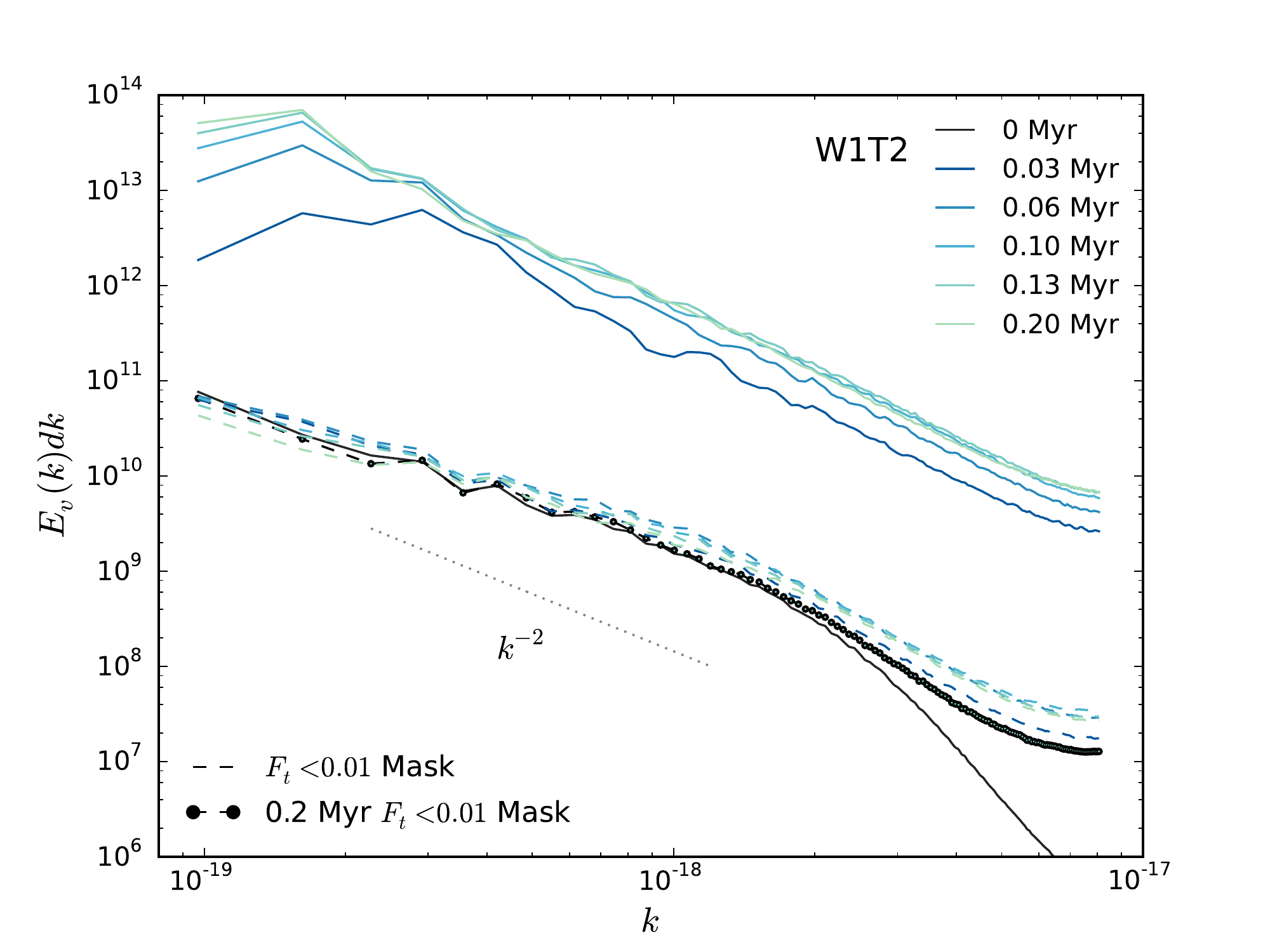}
\end{center}
\caption{Velocity power spectra for the strong-wind run at various times. Solid lines show indicate spectra including all gas, and dashed lines show spectra for gas with $F_t < 0.01$. Dots show the spectral slope at $t=0$ Myr (no winds) when it is masked to include only gas at locations in the $t=0.2$ Myr output with $F_t<0.01$.  At $t=0$ Myr, the fitted spectral slope is $-1.54 \pm 0.05$; the slope is $-1.55 \pm 0.03$ when masked to exclude the $t=0.2$ Myr wind region. At $t=0.2$ Myr, the fitted spectral slope is $-1.33 \pm 0.05$ for gas with $F_t< 0.01$ and $-2.14\pm 0.08$ for all gas.
 Winds increase the overall spectrum slope, while the power spectrum of the pristine gas flattens. 
\label{vspec} }
\end{figure}

The power spectra of the gas density, velocity and momenta describe another fundamental characteristic of turbulence\cite{federrath13}. 
The velocity power spectrum is defined as $E_v (k) dk =  \frac{1}{2}\int \hat v \cdot \hat v' 4 \pi k^2 dk$, which produces  $E(k) \propto k^\alpha$. The slope typically ranges from $\alpha = -5/3$ (Kolmogorov) to $\alpha = -2$ (supersonic) turbulence, where strong fields or solenoidally stirred turbulence may also produce flatter slopes\cite{federrath10}.  Figure \ref{vspec} shows the velocity power spectra for the strongest wind run, W1T2, both including and excluding wind material.  We choose W1T2 to illustrate the impact of winds on the power spectra since the strong W1 sources naturally create a significant offset between the wind and non-wind curves. By adding high-velocity material on $\sim 1-3$~pc scales, winds increase the spectral slope, such that  $\alpha < -2$. Figure \ref{vspec} shows that the power spectrum of the pristine gas flattens over time, which suggests the addition of  small-scale modes, i.e., magnetosonic waves.  These curves display a small power excess on intermediate scales compared to the case with no active winds. After 0.2 Myr the slope difference is $\Delta \alpha \sim 0.8$ -- well above the fit uncertainties. The W2 models exhibit similar trends over time. Both strong and weak wind models display similar slopes, which indicates the impact is not a strong function of the wind input for our range in mass-loss rates.

Figure \ref{vspec} also shows the velocity power spectrum for the initial time ($t=0$~Myr) when the gas is masked with the wind distribution at $t=0.2$~Myr. However, the slopes of the total and masked gas are statistically indistinguishable. This illustrates that the slope changes in the gas selected using the wind fraction $F_t< 0.01$ (see Methods) are due to magnetosonic waves and not simply due to the exclusion of part of the domain volume. We note, however, that the masked spectra display an excess at large wavenumbers. This is caused by small-scale modes introduced by Fourier transforming the mask edge. Comparison of masked and unmasked spectra at 0 Myr shows no impact on the fitted slope.   Supplementary Table 2 displays the slope fits for all runs. 

Vector fields can be decomposed into two parts: a divergence-free solenoidal component ($\mathbf{\nabla} \cdot \mathbf{f_s} = 0$), which quantifies the amount of ``stirring" motion, and a curl-free component  ($\mathbf{\nabla} \times \mathbf{f_c} = 0$), which quantifies the amount of compression. Numerical simulations show the evolution of star formation within a cloud is quite sensitive to the relative fractions of solenoidal and compressive motions, where more compressive turbulence produces higher star formation rates\cite{federrath12}. Protostellar outflows produce more stirring than compression, while gravity enhances compression\cite{hansen12,Offner14b,Offner17}. 

\begin{figure}
\begin{center}
\includegraphics[scale=0.55]{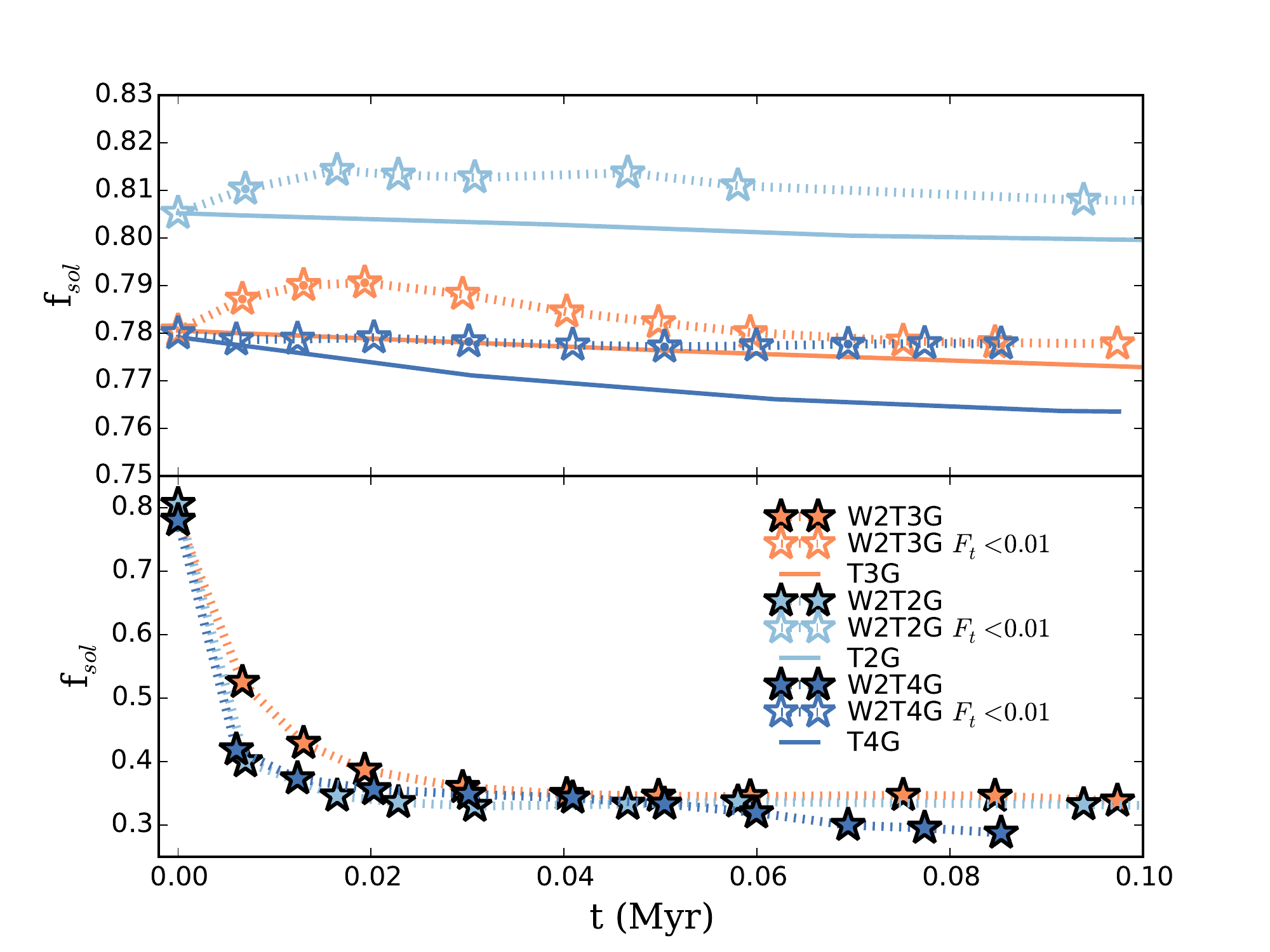}
\end{center}
\caption{Fraction of turbulence that is solenoidal versus time.  Filled stars indicate the solenoidal fraction for all gas including winds, the open stars indicate the fraction for pristine gas ($F_t < 0.01$), and solid lines indicate the solenoidal fraction of models without winds. Gas directly associated with the feedback exhibits more compression, while the solenoidal fraction increases in gas external to the wind shells.
\label{sol} }
\end{figure}

The initial gas turbulence is driven with solenoidal velocity perturbations, which produces a larger fraction of solenoidal motion. Figure \ref{sol}  shows that as the turbulence decays, without the influence of winds the solenoidal fraction decreases as the modes evolve towards equipartition. In contrast, the solenoidal fraction increases in the gas exterior to the wind shells. 
When the wind material is included in the velocity analysis compressive motion dominates due to the strong shock at the shell boundary.

We conclude that winds drive turbulence but do so relatively inefficiently: only $\epsilon \sim 1$\% of the initial wind launching energy is deposited in the cloud outside the shells.  This suggests completely offsetting turbulent dissipation requires non-local input of $\dot E /\epsilon \sim  130 L_\odot$, where the exact value depends on the strength of the magnetic field and turbulent properties. This begs the question:
Is stellar feedback sufficient to maintain turbulence in real molecular clouds?

We consider the Perseus molecular cloud, a low-mass, star-forming region similar in mass and feedback strength to our W2 models, where a number of shells have been identified. The energy injection rate calculated from the shell expansion velocities is  $\dot E\sim 0.5 L_\odot$\cite{arce11}. Meanwhile, the dissipation rate in Perseus is estimated to be $\sim 0.25 L_\odot$ based on mean cloud properties and a dissipation rate coefficient adopted from numerical simulations\cite{arce11}. Although these estimates are each uncertain by a factor of 2-3 \cite{arce11}, they are comparable, and prior work concluded the total feedback energy is sufficient to maintain turbulence in Perseus\cite{arce11}. This estimate, however, assumes the shell expansion itself constitutes turbulent energy, and we argue it, instead, provides relatively local energy input.  
However, we find a significant amount of energy is carried away and deposited outside the shells by magnetosonic waves. The estimated cloud dissipation rate for Perseus is similar to the energy deposited outside the shells due to non-local driving in our W2 simulations: $0.2-0.4 L_\odot$.  Excitation of magnetosonic waves, alone, appears sufficient to offset a large fraction 
of the turbulent dissipation. 
Although the uncertainties in the turbulent dissipation rate, source properties, cloud mass and magnetic field are large, this constitutes remarkable agreement and suggests feedback ``action at a distance" may replenish turbulence and slow gravitational collapse in star-forming clouds.


\begin{methods}\label{methods}

\section{Equations and Algorithms.} 

We perform the calculations using {\sc ORION2}, a parallel, adaptive mesh refinement (AMR) code.  {\sc ORION2} solves the equations for a magnetized compressible gas using a conservative second order Godunov scheme\cite{li15}. Self-gravity is included through a coupled multi-grid method, which solves the Poisson equation\cite{truelove98,klein99}. We treat the stellar wind sources using Lagrangian point particles\cite{krumholz04,Offner15}. The full set of equations solved are:
\begin{eqnarray}
\frac{\p\rho}{\p t} &=& - \nabla\cdot(\rho\mathbf{v}) + \sum_i \dot{M}_{w,i} W_w(\mathbf{x}-\mathbf{x}_i)  \label{eqn:continuity} \\
\frac{\p (\rho\mathbf{v})}{\p t} &=& -\nabla\cdot(\rho\mathbf{v}\mathbf{v} - \frac{1}{4\pi}\mathbf{B}\mathbf{B}) - \nabla P - \rho \nabla \phi + \sum_i \dot{\mathbf{p}}_{w,i} W_w(\mathbf{x}-\mathbf{x}_i) \label{eqn:euler} \\ 
\frac{\p\rho e}{\p t} &=& -\nabla \cdot \left[ \left(\rho e+P\right)\mathbf{v} - \frac{1}{4\pi} \mathbf{B}(\mathbf{v}\cdot\mathbf{B}) \right] - \rho\mathbf{v}\cdot\nabla(\phi) + \dot{\mathcal{E}}_{w,i} W(\mathbf{x}-\mathbf{x}_i) \label{eqn:energy} \\
\frac{\p \rho_t}{\p t} &=& - \nabla\cdot(\rho_t\mathbf{v}) + \sum_i \dot{M}_{w,i} W_w(\mathbf{x}-\mathbf{x}_i)  \label{eqn:tcontinuity} \\
&&\frac{d}{dt}M_i =  - \dot M_{w,i} \\
&& \frac{d}{dt}\mathbf{x}_i = \frac{\mathbf{p}_i }{M_i}\\
&& \frac{d}{dt}\mathbf{p}_i = -{M_{w,i}}\nabla \phi - \dot \mathbf{p}_i \\
\nabla^2 \phi& = & 4 \pi G \rho + 4 \pi G \sum_i M_i \delta (\mathbf{x} - \mathbf{x_i}) \\
\frac{\p \mathbf{B}}{\p t} &=& -\nabla \cdot (\mathbf{v} \mathbf{B} - \mathbf{B}\mathbf{v}) \label{eqn:induction}
\end{eqnarray}
Equations \ref{eqn:continuity}-\ref{eqn:energy} describe conservation of mass density ($\rho$), momentum density ($\rho \mathbf{v}$) and energy density ($e$), where $P$, $\mathbf{v}$, $\mathbf{B}$ and $\phi$ are the gas pressure, velocity, magnetic field and gravitational potential, respectively. 
Equation \ref{eqn:tcontinuity} follows the evolution of $\rho_w$, a passively advected scalar quantity that represents the mass density of the wind material.
We treat the gas as an effectively isothermal fluid such that $e=(1/2) \mathbf{v}^2+P/[\rho(\gamma-1)]$ and the ratio of specific heats is $\gamma=1.0001$.  
$M_i$, $\mathbf{x_i}$ and $\mathbf{p_i}$ are the mass, position and momentum of the $i$th star particle. 

Wind launching occurs in a radius of eight cells centered on the source. Mass, momentum and energy are deposited into these cells at the rates $\dot M_{w,i}$, $\dot \mathbf{p}_{w,i}$ and $\dot{\mathcal{E}}_{w,i}$, respectively,  according to a weighting kernel,  $W$, in which the material is distributed isotropically\cite{cunningham11,Offner15}. The wind prescription follows a mass-loss model for main sequence stars\cite{vink01}, where the wind is initialized to $10^4$~K.  The local temperature effectively reflects the mass-weighted average of the hot wind and cold cloud gas. In the shell, which is mostly swept up 10~K gas,  the temperatures are $\sim$20-100~K, depending on the fraction of wind material.
We do not include cooling, because dynamically, we expect thermal pressure to be sub-dominant. The ratio of the thermal pressure to the ram pressure is generally a few percent at most. Furthermore, the cooling time of gas in the wind cavity exceeds the simulation run time.  

The strongest wind sources will also emit ionizing radiation\cite{gendelev12,rosen16}. However, despite the observed high mass-loss rates, the masses of sources associated with bubbles in local star-forming regions (they are B-type and later) do not produce significant ionizing flux, so we neglect ionization here. Synthetic observations of the simulated sources show good agreement with intermediate-mass stars observed both in Perseus and regions along the Galactic plane\cite{xu17}.

We exploit the AMR capability to add additional resolution in high-density regions and locations where large density jumps are present. We insert cells a factor of 2 smaller,  when the density exceeds the effective Jeans density, which is corrected fro the magnetic pressure:
\begin{equation}
\rho_J = J^2 \frac{\pi k_{\rm B} T}{\mu m_{\rm H} G \Delta x^2} \left( 1+ \frac{0.74}{\beta}\right),
\end{equation}
where $\Delta x$ is the cell size, and we adopt $J=0.125$\cite{truelove97}. Refinement is also added when the density gradient between cells exceeds 10, i.e., $\Delta \rho / \rho \ge 10$. This ensures the wind shocks are resolved on the finest level. 

\section{Initialization.} 

All simulations begin with a uniform density and uniform magnetic field in the  $z$ direction,  except W2T0, which has $\mathbf{B}=0$. We adopt periodic boundary conditions and initialize the turbulence by perturbing the gas with random motions that have power distributed between wavenumbers $k\sim1-2$. This represents energy injected on the scale of the simulation domain ($L-L/2$), i.e., driving from the ISM. The fiducial domain basegrid is 256$^3$, and we allow 2 AMR levels.  After two Mach crossing times the turbulent cascade reaches a statistical steady state. We then place five stellar sources with randomly drawn positions and masses. These sources represent main sequence stars that have finished accreting or have wandered from a nearby cloud. Consequently, their locations are not correlated with the dense gas. One set of sources represents typical mass-loss rates measured in the Perseus molecular cloud (W2),   while the other set includes more massive stars more typical of the Orion high-mass star-forming region (W1).  The total mass-loss rates for W2 and W1 are $4.5 \times 10^{-6} M_\odot$yr$^{-1}$ and $4.2 \times 10^{-5} M_\odot$yr$^{-1}$, respectively \cite{Offner15}, and the individual sources span $2\times 10^{-8}-2\times10^{-5}~\msun$yr$^{-1}$.

\section{Characterizing the Wind Impact.} 

For each grid cell, {\sc ORION2}  stores the mass density, $\rho_{\rm w}$, that was launched by the wind model in a tracer field that is advected with the flow. 
We define $F_t = \rho_{\rm w}/\rho_{\rm total}$ as the fraction of mass in a given cell that originated in a stellar wind.
To separate cells with and without wind material, we set a critical cutoff fraction below which gas can be considered ``pristine".  Supplementary Figure 1 shows cuts through the density and wind fraction centered on a stellar source. Over time, the wind creates a large evacuated bubble. Inside, $F_t \simeq 1$, i.e., all the interior gas is wind material. On the boundary of this region, the wind sweeps up ambient material and the wind fraction drops to  $\sim 0.1$. Thereafter, the amount of wind material rapidly falls below $F_t =$0.01, which we adopt as our fiducial cutoff.  At 0.1 Myr $\sim 90$\% and $98$\% of the volume has $F_t <0.01$ in W1 and W2, respectively.

{\sc ORION2} conserves total mass density, but numerical precision limits mean that $F_t$ is never identically zero.  Thus, it is necessary to adopt a maximum value of $F_t$ to define non-wind material.  Supplementary Figure 2 shows the velocity dispersion as a function of time for three different values of $F_t$ for runs W2B1 and W2B01. These runs contain no turbulence and have constant field strengths. Their Alfv\'en velocities are 0.08~kms$^{-1}$ and 2.7~km s$^{-1}$, respectively. In the weak-field run (W2B1), the magnetic wave does not propagate far beyond the expanding shell. Consequently, the dispersion of the gas with $F_t<10^{-3}$ and $F_t<10^{-4}$ are noticeably lower. In contrast, in the strong-field case (W2B01), which has a $\beta$ similar to our turbulent calculations, the waves propagate further beyond the expanding shell and all three cutoffs return similar dispersions. Supplementary Figure 2 also demonstrates that the net wind impact is slightly smaller in the strong-field case, as expected for shells expanding in a highly pressurized medium\cite{koo92}. Despite this, at most times the external velocity dispersion of the pristine gas is comparable to or higher than the dispersion in the weak field case, a sign of the significance of the excited waves. 

We also check the impact of the choice of $F_t$ on the solenoidal fraction in the turbulent runs.  Adopting $F_t = 10^{-3}$, reduces the solenoidal fraction of the non-wind gas by $<0.5$\%, a negligible difference.

The similarity of the power spectrum slopes of the pristine gas in the strong and weak wind calculations gives additional confidence that the shocks directly produced by the winds are excluded by the fiducial cutoff. Our fiducial value also eliminates gas that would be considered to be part of the wind by an observer, i.e., all of the swept-shell.

\section{Resolution and Numerical Limitations.} 

To check convergence we run a simulation with 512$^3$ basegrid resolution and 2 AMR levels (W2B01HR). Supplementary Figure 3 shows the velocity dispersion as a function of time for W2B01HR compared to an otherwise identical lower resolution run (W2B01). The evolution remains similar, and the higher resolution velocity dispersion is within a few percent of that of the fiducial resolution.

For the strongest sources, some grid imprinting occurs along the cardinal directions (e.g., top-left panel in the movie), which is one weakness of performing spherically symmetric problems on a Cartesian mesh. Offner \& Arce show that the wind-evolution agrees well with the expected analytic solution.  For sources embedded in already turbulent gas the appearance of grid imprinting is minimal.




\end{methods}

{\bf Supplementary Information} is available in the online version of the paper.

{\bf Acknowledgements} S.S.R.O. thanks 
A.~Lee, B.~Gaches and P.~Kumar for helpful comments. S.S.R.O. acknowledges support from NSF Career grant AST-1650486.  The data analysis, images and animations were made possible by {\it yt}, an open-source python package for analyzing and visualizing volumetric data. 
Some of the simulations were performed on the Yale University Omega cluster, which is supported in part by the facilities and staff of the Yale University Faculty of Arts and Sciences High Performance Computing Center. The rest of the simulations were carried out on resources at the Massachusetts Green High-Performance Computing Center, which are supported by staff at the University of Massachusetts.

{\bf Author Contributions } S.S.R.O. performed all the simulations without self-gravity, carried out the analysis, produced the figures and wrote the paper. Y. L. carried out the simulations with gravity and performed a preliminary analysis.

{\bf Competing Interests}  The authors declare that they have no competing financial interests.

{\bf Data Availability}  The data that support the plots within this paper and other findings of this study are available from the corresponding author upon reasonable request.

{\bf Author Information} Reprints and Permissions information is available at npg.nature.com/reprintsandpermissions. The authors declare they have no competing financial interests. Correspondence and requests for materials should be addressed to S.S.R.O. (soffner@astro.as.utexas.edu).

\newpage

\section*{Supplementary material}

Supplementary Table 1: Model Properties

| Name | $\beta$ | Winds ($10^{-6} M_\odot \text{yr}^{-1}$) | Gravity |
|---|---|---|---|
| T0 | $\infty$ | ... | ... |
| T2 | 0.02 | ... | ... |
| T4 | 0.01 | ... | ... |
| T2G | 0.02 | ... | ✓ |
| T3G | 0.03 | ... | ✓ |
| T4G | 0.01 | ... | ✓ |
| W2B1 | 0.1 | 4.5 | ... |
| W2B01 | 0.01 | 4.5 | ... |
| W2B01HR | 0.01 | 4.5 | ... |
| W1T2 | 0.02 | 41.7 | ... |
| W2T0 | $\infty$ | ... | ... |
| W2T2 | 0.02 | 4.5 | ... |
| W2T4 | 0.01 | 4.5 | ... |
| W2T2G | 0.02 | 4.5 | ✓ |
| W2T3G | 0.03 | 4.5 | ✓ |
| W2T4G | 0.01 | 4.5 | ✓ |

Model name, mass-weighted ratio of thermal to magnetic pressure at the end of the driving phase ($\beta = \langle 8\pi \rho c_s^2 / [\bar{B}_x^2 + \bar{B}_y^2 + \bar{B}_z^2] \rangle$), total mass-loss rate due to winds and whether self-gravity ("G") is included. "T" denotes a turbulent environment and "B" a run with uniform field and no turbulence. Model W2B01HR adopts a $512^3$ basegrid in addition to 2 AMR levels.

Supplementary Table 2: Turbulence Properties

| Name | $\dot{E}$ ($L_\odot$) | $\alpha$ | $\alpha_{0.01}$ | $f_{\text{sol}}$ | $f_{\text{sol},0.01}$ |
|---|---|---|---|---|---|
| T0 | -1.0 | $-2.15 \pm 0.05$ | | 0.64 | |
| T2 | -1.5 | $-1.57 \pm 0.05$ | | 0.81 | |
| T4 | -1.2 | $-1.78 \pm 0.09$ | | 0.78 | |
| T2G | -1.5 | $-1.57 \pm 0.05$ | | 0.80 | |
| T3G | -1.1 | $-1.44 \pm 0.03$ | | 0.77 | |
| T4G | -1.3 | $-1.79 \pm 0.08$ | | 0.76 | |
| W1T2 | -1.1 | $-2.28 \pm 0.06$ | $-1.44 \pm 0.04$ | 0.37 | 0.75 |
| W2T0 | -1.0 | $-2.20 \pm 0.08$ | $-2.05 \pm 0.06$ | 0.41 | 0.66 |
| W2T2 | -1.3 | $-1.9 \pm 0.1$ | $-1.46 \pm 0.06$ | 0.52 | 0.80 |
| W2T4 | -1.3 | $-2.6 \pm 0.2$ | $-1.73 \pm 0.08$ | 0.27 | 0.79 |
| W2T2G | -1.3 | $-2.6 \pm 0.2$ | $-1.45 \pm 0.05$ | 0.33 | 0.81 |
| W2T3G | -1.0 | $-2.4 \pm 0.1$ | $-1.44 \pm 0.04$ | 0.34 | 0.78 |
| W2T4G | -1.1 | $-2.5 \pm 0.2$ | $-1.79 \pm 0.07$ | 0.28 | 0.78 |

Model name, the turbulence decay rate, velocity power spectrum slope and solenoidal fraction for gas with $F_t < 0.01$ at $t = 0.1$ Myr.

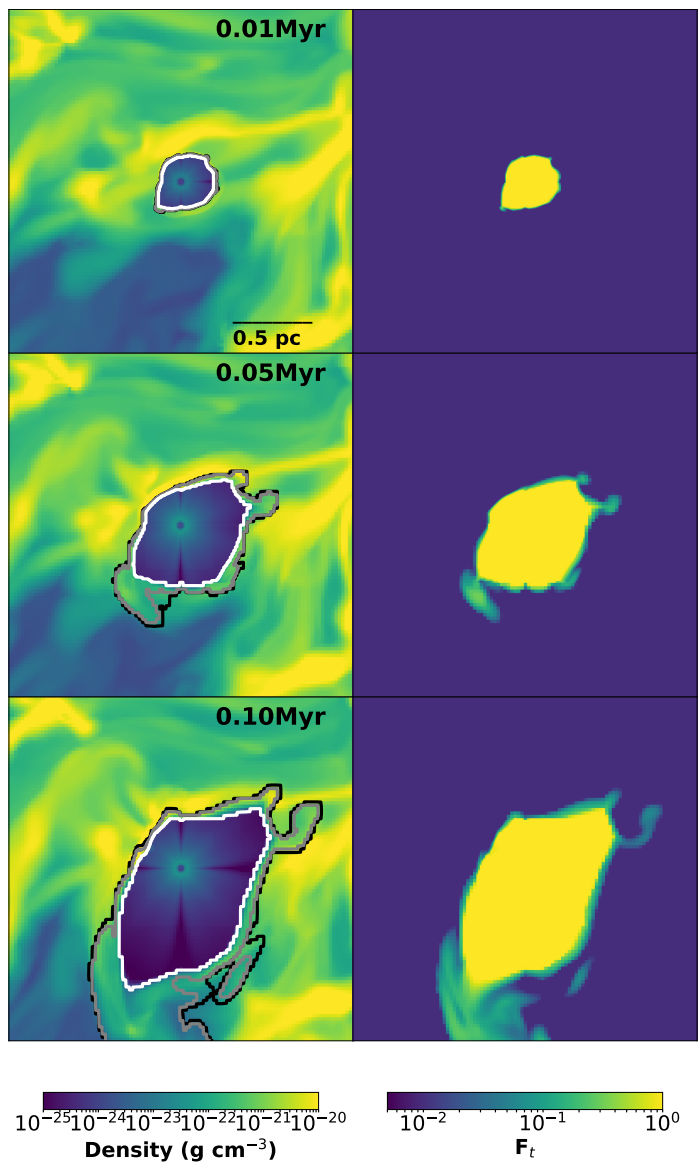

Supplementary Figure 1: Slices through the gas density (left) and wind fraction, $F_t$, (right) centered on a source with $\dot{M}_w = 9 \times 10^{-7}$ $M_\odot \text{yr}^{-1}$ in W2T2 at three times. On the left contours from white to black indicate $F_t = 0.5, 10^{-2}, 10^{-3}$. On the right, dark purple shows gas with $F_t \leq 0.01$. Each panel is 2 pc on a side.

2

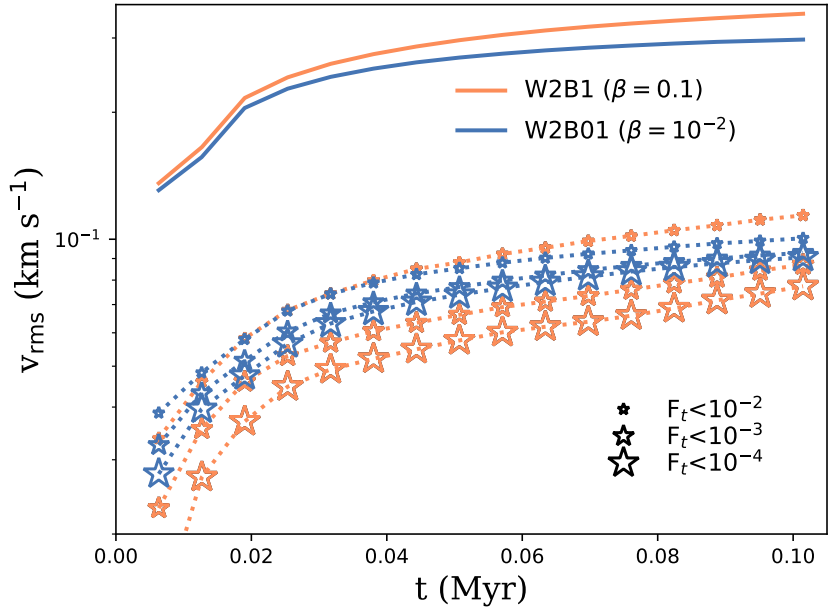

Supplementary Figure 2: Gas velocity dispersion versus time for all gas (solid lines) and gas with $F_t < F_0$ for three values of $F_0$. Both runs begin with a uniform field and no turbulence. The initial field strengths are $B_z = 13.5$ $\mu$G and $B_z = 42.7$ $\mu$G for W2B1 and W2B01, respectively.

3

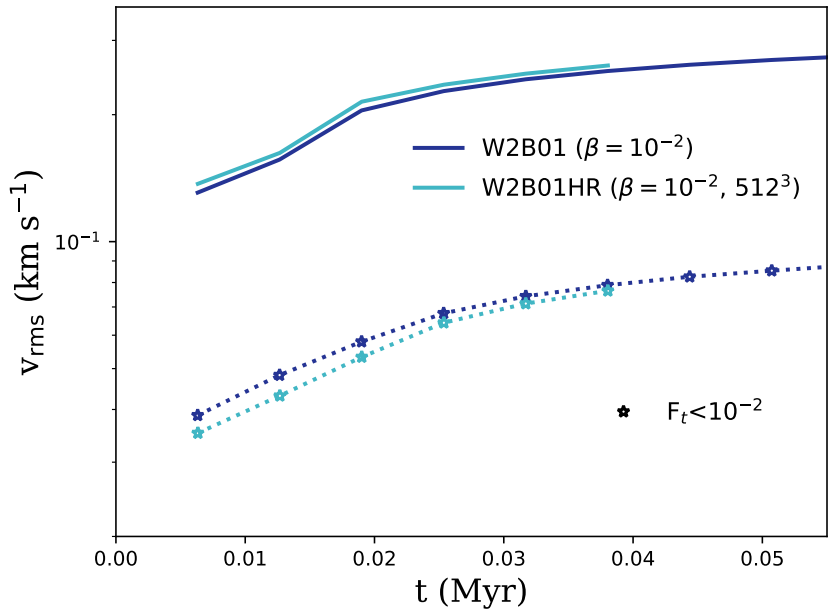

Supplementary Figure 3: Velocity dispersion versus time for a fiducial resolution run, W2B01, and a run with the same conditions and a $512^3$ basegrid, W2B01HR.

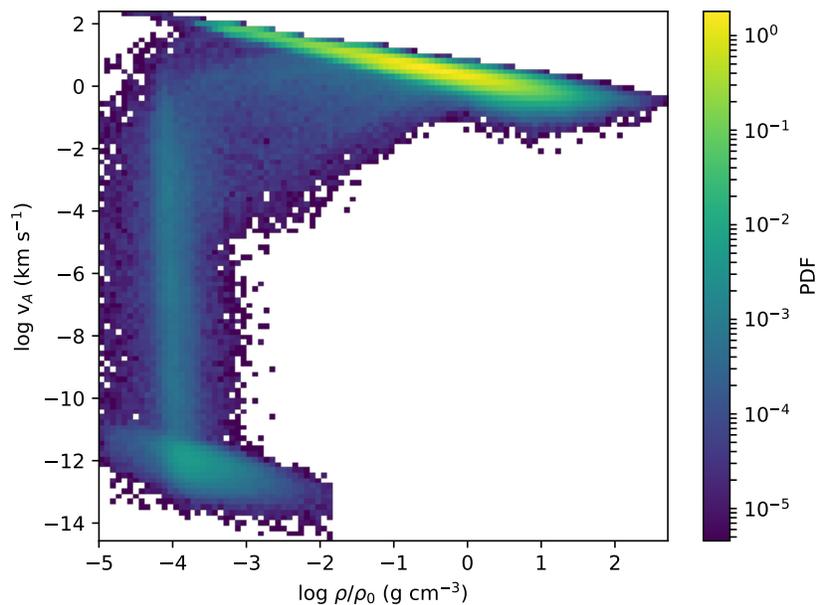

Supplementary Figure 4: The volume fraction of local Alfvén velocities for W2T2 at 0.1 Myr (blue) and 0 Myr (green).

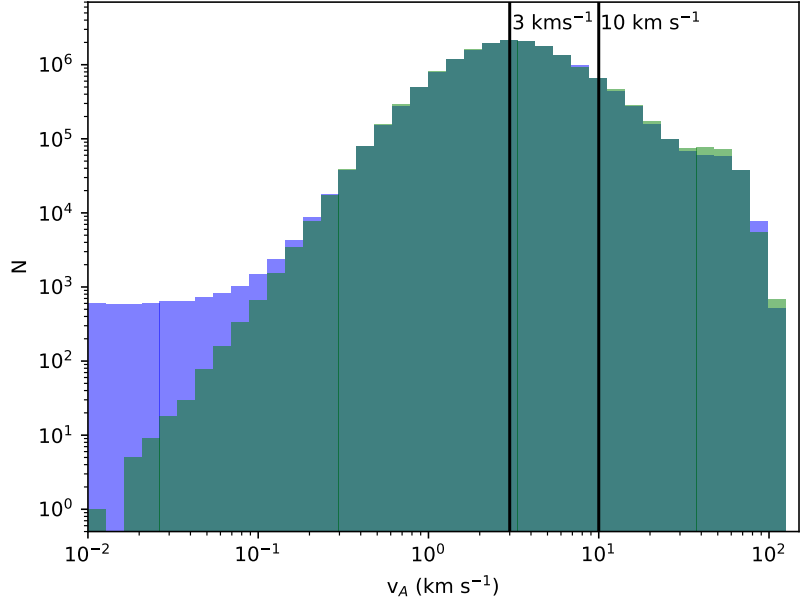

Supplementary Figure 5: The distribution of local Alfvén velocities and densities for W2T2 at 0.1 Myr, where the color scale indicates the volume filling fraction. The densities are normalized by the mean density $\rho_0$.

Supplementary Video 1: The movie shows slices through magnetic field strength ( $|\mathbf{B}| = \left(B_x^2 + B_y^2 + B_z^2\right)^{1/2}$) for two sources in W2T2 (top panels). The arrows indicate the local magnetic field direction. The time is shown in the upper left corner. The mass-loss rates are $9 \times 10^{-7}$ (left) and $1.3 \times 10^{-6}$ $M_\odot \text{yr}^{-1}$(right), respectively. The bottom two panels show the same locations in a run without winds (T2). Waves in $|\mathbf{B}|$ propagate ahead of the bubbles. At later times, the top panels show more structure exterior to the bubbles.






\end{document}